%% file: main.tex
\documentclass{IEEEtran}
\usepackage[english]{babel}
\usepackage[utf8]{inputenc}
\usepackage[T1]{fontenc}
\usepackage{amsmath}
\usepackage{amssymb}
\usepackage[hashEnumerators, smartEllipses]{markdown}

\usepackage{parskip}
\usepackage[T1]{fontenc}
\usepackage{amsfonts}
\usepackage{graphicx}
\usepackage{algcompatible}
\usepackage{algorithm}
\usepackage{caption} 
\usepackage{subcaption} 

\usepackage{amsthm}   

\usepackage{helvet}
\usepackage{etoolbox}
\usepackage{graphicx}
\usepackage{enumerate}
\usepackage{caption}
\usepackage{booktabs}
\usepackage{xcolor} 
\usepackage[colorlinks, citecolor=cyan]{hyperref}
\usepackage{scrextend}
\usepackage{fancyhdr}
\usepackage{amsmath}
\usepackage{amsfonts}
\usepackage{enumitem}
\usepackage{balance}
\usepackage{multirow}
\usepackage{multicol}
\usepackage{enumerate}
\usepackage{float}
\usepackage{ulem}
\usepackage{algpseudocode}
\theoremstyle{definition}

\usepackage{enumerate}
\algrenewcommand\textproc{}

\usepackage{algpseudocode,algorithmicx}
\usepackage{xcolor}
\floatname{algorithm}{Algorithm}
\algnewcommand{\algorithmicand}{\textbf{ and }}
\algnewcommand{\algorithmicor}{\textbf{ or }}
\algnewcommand{\algorithmicnot}{\textbf{ not }}
\algnewcommand{\algorithmicfalse}{\textbf{ false }}
\algnewcommand{\algorithmictrue}{\textbf{ true }}
\algnewcommand{\algorithmicbreak}{\textbf{break}}
\algnewcommand{\OR}{\algorithmicor}
\algnewcommand{\AND}{\algorithmicand}
\algnewcommand{\NOT}{\algorithmicnot}
\algnewcommand{\FALSE}{\algorithmicfalse}
\algnewcommand{\TRUE}{\algorithmictrue}
\algnewcommand{\Break}{\algorithmicbreak}
\newtheorem{thm}{{Theorem}}[section]

\newtheorem{defn}[thm]{Definition}

\newcommand{\compresslist}{ 
	\setlength{\itemsep}{1pt}
	\setlength{\parskip}{0pt}
	\setlength{\parsep}{0pt}
}
\algnewcommand\algorithmicclass{\textbf{class}}
\algdef{SE}[CLASS]{Class}{EndClass}[1]{\algorithmicclass\ \textproc{#1}}{\algorithmicend\ \algorithmicclass}

\algnewcommand\algorithmicenum{\textbf{enum}}
\algdef{SE}[ENUM]{Enum}{EndEnum}[1]{\algorithmicenum\ \textproc{#1}}{\algorithmicend\ \algorithmicenum}

\algnewcommand\algorithmicwhen{\textbf{when}}
\algdef{SE}[WHEN]{When}{EndWhen}[1]{\algorithmicwhen\ #1 \algorithmicdo}{\algorithmicend\ \algorithmicwhen}

\algnewcommand\algorithmicconstructor{\textbf{constructor}}
\algdef{SE}[CONSTRUCTOR]{Constructor}{EndConstructor}[1]{\algorithmicconstructor\ (#1)}{\algorithmicend\ \algorithmicconstructor}
\begin{document}
\title{Enhanced Pruning for Distributed Closeness Centrality under  Multi-Packet Messaging}
%
%
\author{
Patrick D. Manya$^*$, Eugene M. Mbuyi, Gothy T. Ngoie and Jordan F. Masakuna\\
\IEEEauthorblockA{University of Kinshasa, Kinshasa, DR. Congo\\}
\thanks{Corresponding author: patrickmanya@gmail.com}}


\markboth{}%
{Shell \MakeLowercase{\textit{et al.}}: Bare Demo of IEEEtran.cls for IEEE Journals}
\maketitle              
\begin{abstract}
Identifying central nodes using closeness centrality is a critical task in analyzing large-scale complex networks, yet its decentralized computation remains challenging due to high communication overhead. Existing distributed approximation techniques, such as pruning, often fail to fully mitigate the cost of exchanging numerous data packets in large network settings.
In this paper, we introduce a novel enhancement to the distributed pruning method specifically designed to overcome this communication bottleneck. Our core contribution is a technique that leverages multi-packet messaging, allowing nodes to batch and transmit larger, consolidated data blocks. This approach significantly reduces the number of exchanged messages and minimizes data loss without compromising the accuracy of the centrality estimates.
We demonstrate that our multi-packet approach substantially outperforms the original pruning technique in both message efficiency (fewer overall messages) and computation time, preserving the core approximation properties of the baseline method. While we observe a manageable trade-off in increased per-node memory usage and local overhead, our findings show that this is outweighed by the gains in communication efficiency, particularly for very large networks and complex packet structures. Our work offers a more scalable and efficient solution for decentralized closeness centrality computation, promising a significant step forward for large-scale network analysis.
\end{abstract}

\input{introduction}

\input{literature}
\input{description}

\input{results}

\input{conclusion}

\bibliographystyle{splncs04}
 

 
\balance
\bibliography{references}

%
%
%

\end{document}

%% file: introduction.tex
\section{Introduction}\label{sec:introduction}
Centrality metrics are fundamental to network analysis \cite{lam1979congestion}. For measures like betweenness centrality, accurate assessment often requires comprehensive network knowledge, which is computationally prohibitive in large-scale graphs. While complete centrality information is unnecessary for certain distributed tasks (e.g., leader election), the relative centrality of nodes is critical for applications such as active sensing \cite{nelson2006sensory}, robotics coordination \cite{masakuna2019coordinatedextension,jmf2020} and minimizing  synchronization time \cite{ramirez1979distributed,kim2013leader,masakuna2019coordinated}. We specifically focus on closeness centrality and the challenge of efficiently identifying the most central nodes in a decentralized manner.

In large-scale decentralized systems, not all nodes need to maintain a global view of the network; many nodes can quickly determine they possess low closeness centrality and be excluded from further complex computation. The challenge is formalizing and identifying such ``prunable'' nodes within a decentralized algorithm. This was recently addressed by Masakuna et al. \cite{masakuna2023distributed} with their pruning method, which efficiently approximates closeness centrality. We aim to  extend the applicability and efficiency of this baseline method.

This work proposes modifications to the decentralized pruning method to robustly compute node closeness centrality, particularly in environments dominated by high-volume data exchange. Since we employ an approximate and decentralized construction of a communication graph view, nodes are expected to converge on different effective network topologies.

Our proposed enhancement is applicable across any distributed network, offering substantial benefits in very large networks where minimizing message count is paramount. Relevant applications include instrumented vehicles, monitoring systems, and mobile ad-hoc networks. Specifically, we focus on integrating multi-packet messaging \cite{shahid2021concurrent} into the pruning approach. The original method relies on traditional single-packet messaging, which is prone to issues like congestion \cite{schneider2016practical} and packet loss \cite{borella1998internet}, especially in bandwidth-constrained or high-latency environments \cite{kleinrock1992latency}. Introducing multi-packet communication, while necessary for transmitting large data volumes, risks a significant increase in the total number of messages shared. Therefore, a specialized enhancement to the pruning mechanism is required to mitigate the associated communication burden. 

\textbf{Contributions.}
\label{sec:network_contribution}
We make the following contributions toward efficient, distributed identification of closeness central:

\begin{itemize}
\item[(a)] We formally extend the distributed pruning method \cite{masakuna2023distributed} to operate robustly in scenarios utilizing multi-packet messaging. This addresses the critical issue of communication overhead in high-throughput, large-scale networks.

\item[(b)]  We provide a comprehensive analysis of the communication overhead inherent to multi-packet messaging in decentralized settings, specifically investigating its impact on data loss and overall running time. We show that while multi-packet messaging introduces a trade-off in increased memory and local overhead, our proposed enhancements mitigate these issues, preserving the core advantages of the baseline method.

\item[(c)]  We introduce a novel communication overload limiting approach for prunable nodes. This technique ensures that centrality estimation accuracy is maintained while simultaneously minimizing the message exchange burden placed on non-central nodes.
\end{itemize}

We adopt the standard assumptions from \cite{masakuna2023distributed}:
\begin{itemize}
	\compresslist
	\item communication is FIFO, bidirectional, and asynchronous;
	\item each node maintains its own round counter;
	\item nodes are uniquely identifiable;
	\item a node knows the identifiers of its neighbors.
\end{itemize}

%% file: literature.tex
\section{Related work}
The pruning technique, as previously proposed, seeks to optimize the view construction process in distributed networks by allowing certain nodes to cease relaying neighbor information. Consequently, these pruned nodes do not participate in subsequent algorithmic steps, with their closeness centralities effectively treated as zero.
\subsection{Closeness centrality}
Let $\sigma_{ij}$ denote the path distance between the nodes $v_i$ and $v_j$ in an (unweighted) graph $G$\label{you:G} with vertex set $\mathcal{V}$ and edge set $\mathcal{E}$. The path distance between two nodes is the length of the shortest path between these nodes. 
\begin{defn}
	The closeness centrality \cite{bavelas1950communication} of a node is the reciprocal of the average path distance from the node to all other nodes. 
	Mathematically, the closeness centrality $c_i$\label{you:ci} is 
	\begin{equation}
	\label{eq:closeness}
	c_i =  \frac{|\mathcal{V}|-1}{\sum_{j} \sigma_{ij}}\,.
	\end{equation}
\end{defn}	
Nodes with high closeness centrality score have short average path distances to all other nodes.

Closeness centrality has several applications, including social network analysis \cite{freeman2002centrality}, transportation and logistics, and epidemiology \cite{valente2010social}. For example, closeness centrality is used to identify influential individuals in social networks. By measuring how quickly one can reach others in the network, researchers can determine key players or spreaders of information.

\subsection{Other centrality metrics}
Some other network centrality metrics commonly used for network analysis \cite{landherr2010critical}:
\begin{itemize}
    \item \textbf{Betweenness centrality}. It measures the extent to which a node lies on parts between other nodes \cite{newman2005measure}. Nodes with high betweenness centrality act as bridges between different parts of the network, potentially connecting different node communities.
      \item \textbf{Degree centrality}. It defines the number of direct  (one-hop) neighnours a node has \cite{maharani2014degree}. Nodes with high degree centrality are hubs and may be part of densely connected communities.
       \item \textbf{Eigenvector centrality}. A node in a graph is more central if it is connected to other important nodes \cite{newman2005measure}. Nodes with high eigenvector centrality are well-connected to other well-connected nodes and may be part of influential communities.
        \item \textbf{PageRank}. It is a variant of Eigenvector centrality designed for ranking web content using hyperlinks between pages as a measure of importance \cite{zhang2022pagerank}. Nodes with high PageRank are influential and may be part of significant communities.
\end{itemize}
Different centrality metrics provide different perspectives on node importance and connectivity. 
Here, we focus on closeness centrality because the original work was exclusively on it.

\subsection{Original pruning strategy}

Before detailing their pruning method, the authors assert that it successfully retains information regarding the most central nodes within a graph based on closeness centrality. Their goal was to estimate closeness centrality distributions through pruning, which is closely related to eccentricity centrality \cite{hage1995eccentricity,batool2014towards,meghanathan2015correlation}. The eccentricity of a node, defined as the maximum distance to any other node, is reciprocal to closeness centrality. 
Pruning method is deemed suitable for approximating closeness centralities in graph categories where eccentricity and closeness centrality are highly correlated.

In the original work, pruning occurred after each communication iteration, where nodes assess whether they or their one-hop neighbors could be pruned. This enabled nodes to identify which neighbors had been pruned, reducing unnecessary message exchanges in future iterations and consequently minimizing overall communication \cite{masakuna2023distributed}. Three specific types of nodes were identified for pruning\textemdash leaves, nodes that form triangles and only-receiving nodes (nodes that do not share new information, but only receive).

Each node has its own set of pruned nodes from its vicinity. Notably, the sets of pruned nodes for two directly connected neighbors may differ, and there is no requirement for nodes to share this pruning information.



The authors in \cite{masakuna2023distributed} suggest several future directions for enhancing their work. One research direction is the use of multiple packet messaging. Given the limited bandwidth users often face in practical applications of distributed systems, multi-packet messaging is extremely valuable  \cite{sadjadpour2010capacity,
azketa2012schedulability}.
\subsection{Multi-packet messaging}
The effective transmission of large messages in distributed systems often involves multi-packet messaging techniques, which encompass several key concepts, including fragmentation and reassembly, reliable delivery and flow control.
For example, to accommodate network constraints, large messages are divided into smaller packets (fragments). Each packet is transmitted independently and reassembled at the destination. This method effectively manages bandwidth and reduces the likelihood of transmission failures \cite{postel1981transmission}.
Also,
protocols such as TCP (Transmission Control Protocol) ensure reliable packet delivery \cite{madhuri2016performance}, making them suitable for applications where data integrity is critical. Conversely, protocols like UDP (User Datagram Protocol) prioritize speed over reliability \cite{madhuri2016performance}, which can be advantageous in certain scenarios. Multi-packet systems may incorporate custom acknowledgment mechanisms to verify the receipt of all packets. We expect increased network congestion when pruning large‑scale networks, especially in scenarios where data are transmitted as multiple packets.

To mitigate network congestion, flow control mechanisms are essential for regulating the rate of packet transmission based on the receiver's capacity. Techniques such as sliding window protocols \cite{van1995sliding} enable multiple packets to be in transit before requiring an acknowledgment, enhancing overall throughput \cite{oppenheimer2011top}.
Robust error detection and correction methods, such as checksums \cite{stone1998performance} and redundant packets \cite{felix2018redundant}, are implemented to ensure data integrity. If a packet is lost or corrupted during transmission, only the affected packets are retransmitted rather than the entire message \cite{reed2012error}.

Multi-packet messaging frequently employs adaptive routing strategies, which dynamically select optimal paths for packet transmission based on current network conditions. This adaptability helps reduce latency and improve overall communication efficiency \cite{girard1983dynamic}.
Leveraging multicast and broadcast techniques allows multi-packet messaging to efficiently distribute messages to multiple recipients simultaneously. This capability is particularly beneficial in distributed applications that require data sharing across numerous nodes \cite{jahanshahi2014multicast}.

Multi-packet messaging has found extensive applications in distributed databases, facilitating the efficient transmission of data changes across nodes; in cloud computing, enhancing communication within microservices architectures; and in real-time collaboration tools, ensuring smooth data flow during simultaneous user interactions.
Despite its advantages, multi-packet messaging faces several challenges. Latency can remain an issue due to the time required to transmit multiple packets. Additionally, ensuring that packets arrive in the correct order, especially when taking different paths, is critical. As the number of nodes in a network increases, managing multi-packet communication effectively becomes increasingly complex.

%% file: description.tex
\section{Extended pruning}
\label{sec:description}
To effectively accommodate the pruning method \cite{masakuna2023distributed} when using multi-packet messaging, we explore additional strategies to reduce communication overload.

We structure messages into multiple packets. This design allows each message sent from node $v_i$ to node $v_j$ during the $t$-th iteration to be divided into $m$ packets, denoted as $P_{ij}^{(t)} = \{p_{ij}^{(t,1)}, p_{ij}^{(t,2)}, \ldots, p_{ij}^{(t,m)}\}$. By segmenting messages in this way, we can better manage bandwidth utilization, particularly in networks where message sizes may exceed optimal transmission limits.

Additionally, this packetization reduces the likelihood of message loss in high-density networks, as smaller packets are less prone to being dropped during transmission. Each packet can be sent independently, allowing for more flexible and resilient communication. If some packets are lost, nodes can still function with the remaining packets, ensuring that vital information is not completely compromised. This multi-packet messaging framework not only enhances the efficiency of communication but also supports more robust data integrity during centrality calculations.

To achieve this more effectively, we minimize the number of communications between neighbors. This improved strategy requires each node $v_i$ to send messages to its neighbors a total of $D_i - 2$ times during the communication phase, where $D_i < D$ (with $D$ representing the maximum number of iterations). For instance, in the original method, a leaf node would send messages twice; in this revised approach, it will not send any messages at all. Thus, its direct neighbour will treat it as a leaf. This method reduces communication overload in the system. However, there would be an issue if a prunable node (but not yet pruned by its direct neighbours) fails at the start of the algorithm in which case there will be no way its direct neighbours to be informed of its failure. Here, we assume that nodes do not fail at the start of the algorithm.

Let $s_i = m\times d_i \times D_i$ denote the total number of messages a node $v_i$ sends over $D_i$ iterations using the original pruning method, where $d_i$ denotes its degree. In this revised version, the node $v_i$ will send
\begin{equation}
    \hat{s}_i \leq m\times d_i\times (D_i-2)\,,
\end{equation}
 saving $2\times m\times d_i$ from the initial amount. This saved amount of messages could be high in a very large network.

To achieve this, we employ sliding‑window protocols \cite{van1995sliding} for multi‑packet messaging—particularly the Go‑Back‑N ARQ scheme \cite{yao1995effective}, as we assume a low error rate and require rapid transmission. In contrast, Selective Repeat ARQ \cite{weldon1982improved}, while capable of handling a larger number of errors, introduces additional complexity. The refined method is presented in Algorithm~\ref{algo:pruning}. The notation follows that of the original work \cite{masakuna2023distributed}.

For communication failures, we adopt the neighbour‑coordination strategy used in \cite{sheth2005decentralized}, following the same design principles used in the original pruning method \cite{masakuna2023distributed}. This mechanism enables neighbouring nodes to collaboratively detect missing or delayed messages and to compensate for local communication faults through redundancy and mutual verification. By relying on local consensus rather than global synchronization, the system maintains robustness even under intermittent link failures, ensuring that pruning decisions remain consistent across the network.


\begin{algorithm}
	\begin{algorithmic}[1]
		\scriptsize
		\Procedure{runPruning($\mathcal{N}_i, D$)}{}
  \If{\textcolor{red}{$|\mathcal{N}_i| > 1$}}
		\State \texttt{PruningObject}$\gets$\textsc{pruning}($\mathcal{N}_i, D$)
		\State \texttt{PruningObject}.\textsc{initialOneHop()}
		\State \texttt{PruningObject}.\textsc{initialUpdate()}
		\State \texttt{PruningObject}.\textsc{firstPruningDetection()}
		\While{\NOT \texttt{PruningObject}.\textsc{isEnded()}}
		\State \texttt{PruningObject}.\textsc{nextOneHop()}
		\State \texttt{PruningObject}.\textsc{nextUpdate()}
		\EndWhile
  \EndIf
		\EndProcedure
		\State
		\Class{pruning}
		\State \textbf{\underline{Class variables}}
		\State \makebox[1cm][l]{$D$} the pre-set maximum number of iterations
		\State \makebox[1cm][l]{$\mathcal{F}^{(t)}_{i}$} set of pruned nodes known by $v_i$ at the end of iteration $t$
		\State \makebox[1cm][l]{$\mathcal{F}_{i, t}$} set of pruned nodes known by $v_i$ at the end of iteration $t$
		\State \makebox[1cm][l]{$\mathcal{M}_i$} a message queue for messages received by the node
        \State \makebox[1cm][l]{\textcolor{red}{$P_{ij}^{(t)}$}} \textcolor{red}{a 
  message sent from node $v_i$ to node $v_j$ during the $t$-th iteration}
        \State \makebox[1cm][l]{\textcolor{red}{$p_{ij}^{(t,k)}$}} \textcolor{red}{the $k$th packet of the message $P_{ij}^{(t)}$}
        \State \makebox[1cm][l]{\textcolor{red}{$m$}} \textcolor{red}{the number of packets in a message}
        \State \makebox[1cm][l]{$\mathcal{N}_{i}$} set of immediate neighbours of $v_i$
		\State \makebox[1cm][l]{$\mathcal{N}^\mathrm{up}_{i, t}$} set of neighbours of $v_i$ which are still active up to iteration $t$
		\State \makebox[1cm][l]{$\mathcal{Q}_{ij}$} the one-hop neighbours of node $v_j$ sent to $v_i$ at the first iteration 
		\State \makebox[1cm][l]{$\mathcal{N}^{(t)}_{i}$} set of new nodes discovered by $v_i$ at the end of iteration $t$
		\State \makebox[1cm][l]{$\mathcal{N}_{i, t}$} view of communication graph of $v_i$ up to iteration $t$
		\State \makebox[1cm][l]{$c_i$} the closeness centrality of node $v_i$
		\State \makebox[1cm][l]{$\delta_i$} a variable used for computation of $c_i$
		\State \makebox[1cm][l]{$t$} current iteration number
		\State \makebox[1cm][l]{$T$} the actual maximum number of iterations

		\\\hrulefill
		\Constructor{$\mathcal{N}_i, D$}
		\State $(t, D, \delta_i)\gets (0, D, |\mathcal{N}_{i}|)$
		\State $\mathcal{N}^{(0)}_i\gets  \{v_{j}: \forall v_j\in\mathcal{N}_i \}$
		\State $\mathcal{N}_{i, 0}\gets \mathcal{N}^{(0)}_i$.\texttt{clone()} \Comment{``$v_i$ detects its immediate neighbours''}\label{line:initiation_end}
		
		\EndConstructor
						\State
		\Function{initialOneHop()}{}
		\For{$v_j \in \mathcal{N}_i \textcolor{red}{\;\wedge \;|\mathcal{N}_i|\;=\;1} $}
  		\State \sout{$v_i$ sends $\langle \texttt{NeighbouringMessage($i, \mathcal{N}_i^{(0)}$)}\rangle$
		to $v_j$}
  \State \textcolor{red}{$p_{ij}^{(0,1)},\cdots, p_{ij}^{(0,m)}=\langle \texttt{fragmentation($i, \mathcal{N}_i^{(0)}$)}\rangle$}
  \State \textcolor{red}{\texttt{GoBackNARQ}($\langle p_{ij}^{(0,1)}, p_{ij}^{(0,2)},\cdots, p_{ij}^{(0,m)}\rangle$}

		\EndFor
		\EndFunction
		\State
				\Procedure{initialUpdate()}{}
		\State $\mathcal{N}^{(1)}_i\gets \emptyset$
		\State $t\gets t + 1$
		\While{$\mathcal{M}_i.\texttt{size()}\geq1$}
		\State $(j,\mathcal{N}_j^{(0)}) \gets \textcolor{red}{assembly(}\mathcal{M}_i.\texttt{dequeue}()\textcolor{red}{)}$	
		\State $\mathcal{N}^{(1)}_i\gets \mathcal{N}^{(1)}_i \cup \mathcal{N}^{(0)}_j$
		\Comment{``$v_i$ fuses the messages received''}	
		\State $\mathcal{Q}_{ij}\gets \mathcal{N}^{(0)}_{j}$
		\EndWhile
		\State $\mathcal{N}^{(1)}_i\gets \mathcal{N}^{(1)}_i\setminus \mathcal{N}_{i, 0}$ 
		\State $\mathcal{N}_{i, 1}\gets \mathcal{N}_{i, 0}\cup \mathcal{N}^{(t)}_i$\label{line:first_iteration_end}
		\State $\delta_i\gets \delta_i +t|\mathcal{N}^{(1)}_i|$
		\EndProcedure
				\State
		\Procedure{firstPruningDetection()}{}
		\State \textsc{leavesDetection()} 
		\State  \textsc{triangleDetection()} 
		\State $\mathcal{F}_{i, t}\gets \mathcal{F}^{(t)}_i$.\texttt{clone()}
		\EndProcedure
		\State
		\Procedure{leavesDetection()}{}\Comment{``$v_i$ detects leaves in its neighbourhood}
		\State $\mathcal{F}^{(1)}_i\gets \emptyset$
		\For{$v_j\in \mathcal{N}_i\cup\{v_i\}$}
		\If{$|\mathcal{Q}_{ij}|=1$}

		\algstore{pruningTemplate}
	\end{algorithmic}  
	\caption[\textit{Refined pruning method.}]{\textit{Refined pruning method. Red texts indicate additional information.} \textit{  The algorithm gives the code executed for a single node $v_i$. An example of pseudo code is given in the procedure \textsc{runPruning}.}}
	\label{algo:pruning}
\end{algorithm}
\begin{algorithm}
	\begin{algorithmic}[1]
		\algrestore{pruningTemplate}
		\scriptsize
			\State $\mathcal{F}^{(1)}_i\gets \mathcal{F}^{(1)}_i\cup \{v_j\}$
		\EndIf
		\EndFor
		\EndProcedure
		\State
\Procedure{triangleDetection}{}\Comment{``$v_i$ detects elements causing triangles in its neighbourhood}
\For{$v_j\in \mathcal{N}^\mathrm{up}_{i, 1}\cup\{v_i\}$}
\If{  $|\mathcal{Q}_{ij}|=2$}
\State Let $v_f, v_g$ be the two immediate neighbours of $v_j$
\If{$v_f\in\mathcal{Q}_{ig}$}
\State $\mathcal{F}^{(1)}_i\gets \mathcal{F}^{(1)}_i\cup \{v_j\}$
\EndIf
\EndIf
\EndFor

\EndProcedure
\State
\Procedure{nextOneHop()}{}
\If{\NOT \textsc{isEnded()}}
\State $\mathcal{N}^\mathrm{up}_{i, t}\gets \mathcal{N}^\mathrm{up}_{i, t}\setminus \mathcal{F}_{i, t}$

\For{$v_j \in \mathcal{N}^\mathrm{up}_{i, t}$}
\State \sout{$v_i$ sends $\langle \texttt{NeighbouringMessage($i, \mathcal{N}_i^{(t-1)}$)}\rangle$
to $v_j$}
\State \textcolor{red}{$p_{ij}^{(t,1)},\cdots, p_{ij}^{(t,m)}=\langle \texttt{fragmentation($i, \mathcal{N}_i^{(t-1)}$)}\rangle$}
\State \textcolor{red}{\texttt{GoBackNARQ}($p_{ij}^{(t,1)}, p_{ij}^{(t,2)},\cdots, p_{ij}^{(t,m)}$)}
\EndFor
\EndIf
\EndProcedure
\State
\Procedure{nextUpdate()}{}
\If{\NOT \textsc{isEnded()}}
\State $t\gets t+1$
\State  $\mathcal{N}^{(t)}_i\gets \emptyset$
\While{$\mathcal{M}_i.\texttt{size()}\geq 1$}
\State $(j, \mathcal{N}_j^{(t-1)}) \gets \textcolor{red}{assembly(}\mathcal{M}_i.\texttt{dequeue}()\textcolor{red}{)}$
\State $\mathcal{N}^{(t)}_i\gets \mathcal{N}^{(t)}_i \cup \mathcal{N}^{(t-1)}_j$
\Comment{``$v_i$ fuses the messages received''}
\EndWhile
\State $\mathcal{N}^{(t)}_i\gets \mathcal{N}^{(t)}_i\setminus \mathcal{N}_{i, t-1}$ 
\State $\mathcal{N}_{i, t}\gets \mathcal{N}_{ i, t-1}\cup \mathcal{N}^{(t)}_i$
\State \textsc{furtherPruningDetection()} 
\State $\mathcal{F}_{i, t}\gets \mathcal{F}_{i, t} \cup \mathcal{F}^{(t)}_{i}$
\State $\delta_i\gets \delta_i +t|\mathcal{N}^{(t)}_i|$
\Else
\State $T\gets \min(t, D)$
\State $c_i\gets \textsc{closenessCentrality}(T)$
\EndIf

\EndProcedure
		\State
		\Procedure{furtherPruningDetection()}{}\Comment{``$v_i$ detects elements of $\mathcal{F}^{(t)}_i$ in its neighbourhood''}
		\State $\mathcal{F}^{(t)}_i\gets \emptyset$
		\For{$v_j\in \mathcal{N}^\mathrm{up}_{i, t}$}
		\If{$\mathcal{N}^{(t)}_j\subseteq \mathcal{N}_{i, t-1}$}
		\State $\mathcal{F}^{(t)}_i\gets \mathcal{F}^{(t)}_i\cup \{v_j\}$
		\EndIf
		\EndFor
		\If{$|\mathcal{N}^\mathrm{up}_{i, t}|=1$ \AND $\mathcal{N}_i^{(t)}\neq \emptyset$ }
		\State $\mathcal{F}^{(t)}_i\gets \mathcal{F}^{(t)}_i\cup \{v_i\}$
		\EndIf
		\EndProcedure
		\State

		\Function{closenessCentrality}{$T$}
		\If{$v_i \in \mathcal{F}_{i, T}$}
		\State \Return $ 0$
		\EndIf
		\State \Return $ \frac{|\mathcal{N}_{i, T}|-1}{\delta_i}$
		\EndFunction
		
		\State
		\Function{isEnded()}{}
		\State \Return $t=D$ \OR $\mathcal{N}_i^{(t)}=\emptyset$ \OR $v_i \in \mathcal{F}_{i, t}$
		\EndFunction
		\EndClass
	\end{algorithmic}  
\end{algorithm}

%% file: results.tex
\section{Experimental investigation}
\label{sec:results}
We now investigate on the impact of multi-packet messaging using pruning method.
We use the same configurations as in the original paper \cite{masakuna2023distributed}, with some adjustments and an additional data set.

Our experiments evaluate the following scenarios in order to comprehensively assess the proposed approach: \begin{itemize} \compresslist \item Evaluation of multi‑packet messaging under the original pruning method, focusing on running time and data loss. \item Comparison of the number of messages received by nodes when using the original pruning method versus our enhanced version across a variety of network topologies. \item Comparison of per‑node running times for the original pruning method and our enhancement across different networks. \item Comparison of per‑node memory requirements for the original pruning method and our enhancement across different networks. \item Comparison of the approximated most central nodes produced by our enhancement and by the original pruning method. A good approximation should select a most central node whose distance to the exact most central node is small. \item Statistical validation using the Wilcoxon signed‑rank test \cite{wilcoxon1992individual} and effect size measures \cite{cohen1962statistical} to determine whether the performance differences between our enhancement and the original pruning method are statistically significant. We use a significance level of 
0.01
 for all hypothesis tests. \end{itemize}

\subsection{Experimental setup}
We implemented our method in Python using the \texttt{NetworkX} library \cite{hagberg2013networkx}. To measure memory usage, we used the \texttt{Pympler}\footnote{\texttt{Pympler} is a Python library for measuring, analyzing, and monitoring the memory behavior of objects at runtime.} library, and to record running times of the distributed algorithm, we used Python’s \texttt{Timer}\footnote{\texttt{Timer} is a Python module for handling algorithmic tasks involving time measurement.} module.

We conducted extensive simulations on both randomly generated graphs (constructed as described below) and several real‑world networks.

\subsubsection{Randomly generated networks}

We considered a $250 \times 250$ grid with integer coordinates and generated $100$ random, connected, undirected graphs as follows. We first sampled $N$ grid locations uniformly without replacement to serve as nodes, where $N$ was drawn uniformly from the interval $[100, 2000]$. Two nodes were connected by an edge whenever their Euclidean distance was less than a specified communication range, $d = 10$.  

\begin{figure}[h]
	\centering
	\includegraphics[width=0.45\textwidth]{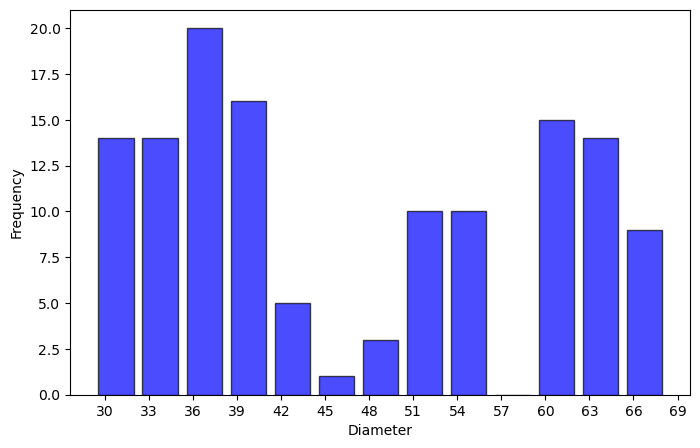}
	\caption[\textit{Diameters of the 135 graphs considered.}]{\textit{Diameters of the 135 graphs considered.}}
	\label{fig:random_property}
\end{figure}

Across these graphs, the number of edges and graph diameters fell within the intervals $[100, 4000]$ and $[30, 68]$, respectively—see Figure~\ref{fig:random_property}. For multi‑packet messaging, the number of packets $m$ was chosen from the set $\{1, 10, 20, 30, 50\}$.

\subsubsection{Real-world networks}

We also evaluated our method on 35 real‑world graphs, including the Facebook artist network \cite{rozemberczki2019gemsec}, a phenomenology collaboration network \cite{leskovec2007graph}, a snapshot of the Gnutella peer‑to‑peer network \cite{leskovec2007graph}, and 32 autonomous system graphs \cite{leskovec2005graphs}.  

The Facebook artist network captures connections between artists on the Facebook platform, reflecting relationships such as direct friendships, shared interests, or group memberships. The phenomenology collaboration network represents co‑authorship links among researchers submitting to the Journal of High Energy Physics. In the Gnutella peer‑to‑peer network, nodes correspond to hosts and edges represent active connections between them. Autonomous system graphs model communication links between Internet routers, derived from Border Gateway Protocol logs.  

Key characteristics of several of these networks are summarized in Table~\ref{tab:autonomous_graph}.
\subsection{Results and discussion}
\label{sec:network_results} 
\subsubsection{Multi-packet messaging}
\begin{figure}[h]
	\centering
		\includegraphics[width=0.45\textwidth]{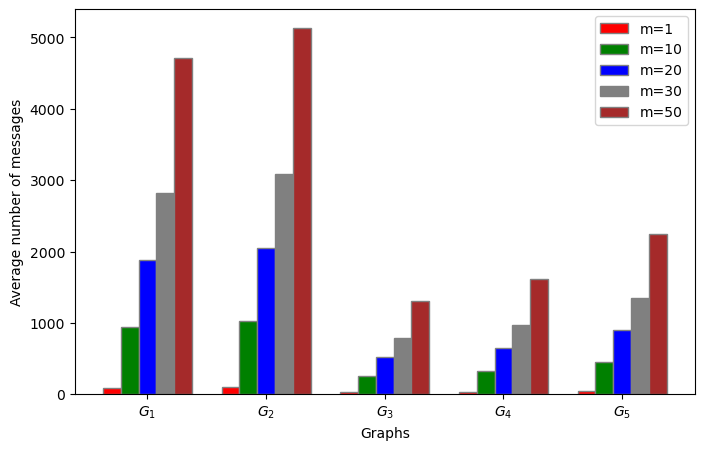}
	\caption[\textit{A sample of average number of messages per node using multi-packer messaging.}]{\textit{{A sample of average number of messages per node using multi-packer messaging.
	}}}
	\label{fig:multipacket}
\end{figure}

Figure \ref{fig:multipacket} shows the average number of messages per node when multi-packet messaging is considered. It indicates that larger amounts of data are transmitted when the number of packets increases, which makes the system more efficient. This expected result has been extensively studied in the community. It has been shown that large amounts of data across a network are efficiently transmitted when multi-packet messaging is applied in distributed systems. A single-packet messaging can lead to issues such as congestion \cite{schneider2016practical} and data loss \cite{borella1998internet}, especially in environments with limited bandwidth or high latency \cite{kleinrock1992latency}. 

\begin{figure}[h]
	\centering
		\includegraphics[width=0.45\textwidth]{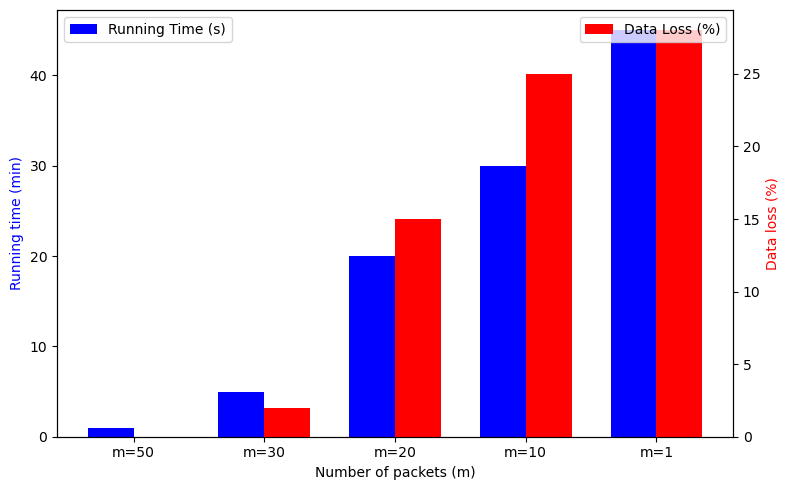}
	\caption[\textit{Running time and data loss.}]{\textit{{Running time and data loss using multi-packer messaging on the original pruning method.
	}}}
	\label{fig:loss}
\end{figure}

Figure \ref{fig:loss} shows running time and data loss when employing multi-packet messaging on the pruning method. It shows that as the number of packets composing a message decreases, the system may experience longer running times and a higher rate of data loss.
Sending large amounts of data on a bandwidth-limited channel can lead to congestion, increased latency, packet loss, and a decrease in effective throughput, which can ultimately affect the performance of applications relying on that data transmission. The analysis indicates that using multi-packet messaging is crucial for minimizing running times and data loss, especially in bandwidth-limited environments.

\begin{figure}[h]
	\centering
		\includegraphics[width=0.45\textwidth]{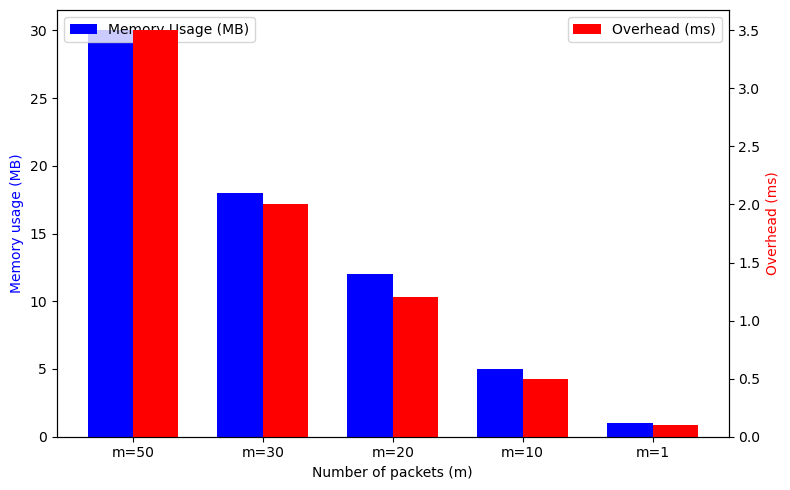}
	\caption[\textit{Memory usage and overhead for multi-packet messaging.}]{\textit{{Memory usage and overhead for multi-packet messaging on the original pruning method.
	}}}
	\label{fig:overhead}
\end{figure}

In scenarios where messages are small, single packet messaging is generally more efficient. But when messages are large, the network can become congested, especially when the amount of data exceeds the capacity of the channel, leading to delays and data loss (as previously shown in Figure \ref{fig:loss}). As the channel gets saturated, the time it takes for packets to travel from the sender to the receiver (latency) increases.
Further, when the data exceeds the channel's capacity, packets may be dropped. This is especially common in TCP/IP networks, where lost packets require retransmission.  For larger messages, multi-packet messaging is necessary, but it will incur greater memory usage and overhead (as shown in Figure \ref{fig:overhead}).

\subsubsection{Average and maximum number of messages}
\begin{figure*}[h]
	\centering	
	\begin{subfigure}[h]{0.325\textwidth}
		\includegraphics[width=\textwidth]{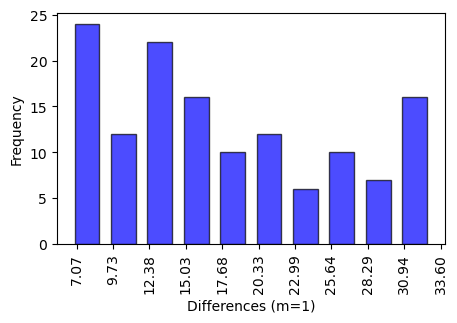}
		\caption{}
		\label{fig:avg_1}
	\end{subfigure}
	\begin{subfigure}[h]{0.325\textwidth}
	\includegraphics[width=\textwidth]{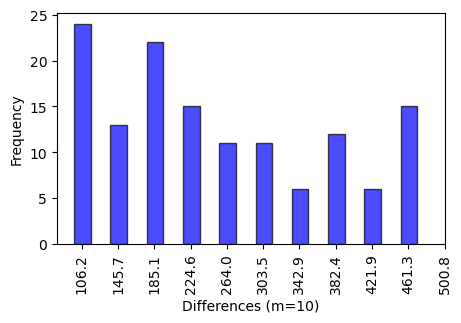}
	\caption{}
	\label{fig:avg_10}
\end{subfigure}
	\begin{subfigure}[h]{0.325\textwidth}
	\includegraphics[width=\textwidth]{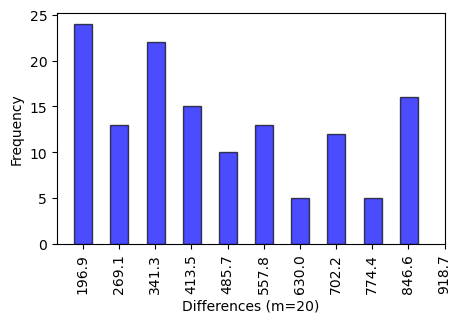}
	\caption{}
	\label{fig:avg_20}
\end{subfigure}
	\begin{subfigure}[h]{0.325\textwidth}
	\includegraphics[width=\textwidth]{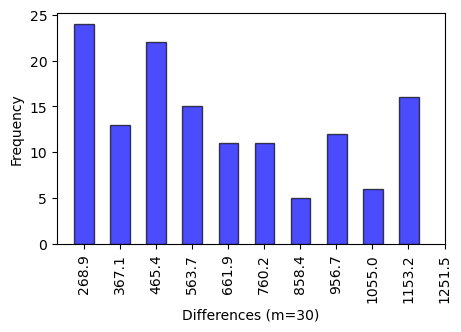}
	\caption{}
	\label{fig:avg_30}
\end{subfigure}
	\begin{subfigure}[h]{0.325\textwidth}
	\includegraphics[width=\textwidth]{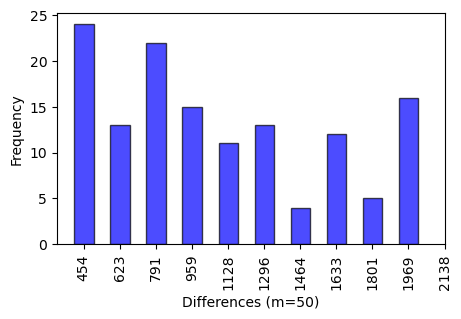}
	\caption{}
	\label{fig:avg_50}
\end{subfigure}
	\caption[\textit{Histograms showing the difference in the average  number of messages on the $135$ graphs.}]{\textit{{Differences in the average number of messages between the original pruning method and our enhancement on the $135$ graphs. Positive values indicate that our enhancement outperfoms the original pruning method.  \textbf{(\ref{fig:avg_1})}: $m=1$.
	\textbf{(\ref{fig:avg_10})}: $m=10$.
 \textbf{(\ref{fig:avg_20})}: $m=20$.
 \textbf{(\ref{fig:avg_30})}: $m=30$.
 \textbf{(\ref{fig:avg_50})}: $m=50$. }}}
	\label{fig:avg}
\end{figure*}

\begin{figure*}[!ht]
	\centering	
	\begin{subfigure}[h]{0.325\textwidth}
		\includegraphics[width=\textwidth]{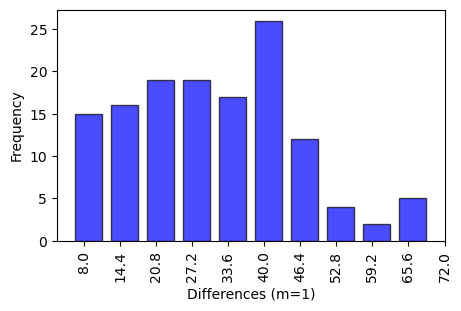}
		\caption{}
		\label{fig:max_1}
	\end{subfigure}
	\begin{subfigure}[h]{0.325\textwidth}
	\includegraphics[width=\textwidth]{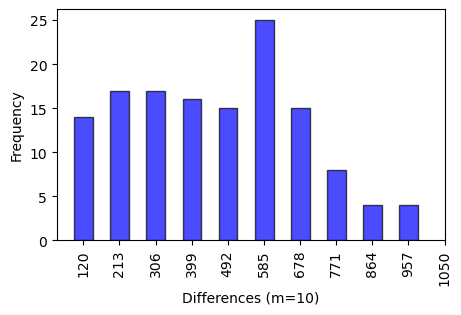}
	\caption{}
	\label{fig:max_10}
\end{subfigure}
	\begin{subfigure}[h]{0.325\textwidth}
	\includegraphics[width=\textwidth]{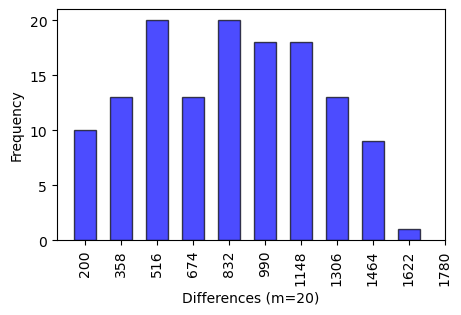}
	\caption{}
	\label{fig:max_20}
\end{subfigure}
	\begin{subfigure}[h]{0.325\textwidth}
	\includegraphics[width=\textwidth]{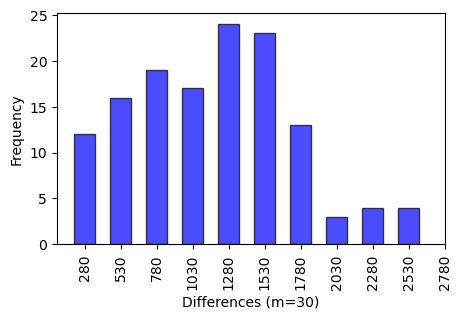}
	\caption{}
	\label{fig:max_30}
\end{subfigure}
	\begin{subfigure}[h]{0.325\textwidth}
	\includegraphics[width=\textwidth]{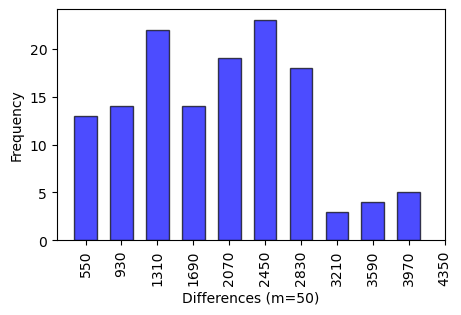}
	\caption{}
	\label{fig:max_50}
\end{subfigure}
	\caption[\textit{Histograms showing the difference in the maximum  number of messages on the $135$ graphs.}]{\textit{{Differences in the maximum number of messages between the original pruning method and our enhancement on the $135$ graphs. Positive values indicate that our enhancement outperfoms the original pruning method.  \textbf{(\ref{fig:max_1})}: $m=1$.
	\textbf{(\ref{fig:max_10})}: $m=10$.
 \textbf{(\ref{fig:max_20})}: $m=20$.
 \textbf{(\ref{fig:max_30})}: $m=30$.
 \textbf{(\ref{fig:max_50})}: $m=50$. }}}
	\label{fig:max}
\end{figure*}
The experiments clearly demonstrate the improved communication performance of our enhancement over the original pruning method \cite{masakuna2023distributed}, achieved through further optimized pruning.
Figures \ref{fig:avg} and \ref{fig:max} explicitly illustrate the reduction in communication load. Figure \ref{fig:avg} shows the differences in the average number of messages per node, and Figure \ref{fig:max} shows the differences in the maximum number of messages per node between the original pruning method and our enhancement, across the entire set of $135$ test graphs.
Quantitatively, our approach successfully reduced the communication load by $5\%$ to $15\%$ on average across all network topologies tested.

\begin{figure*}[!ht]
	\centering	
	\begin{subfigure}[h]{0.325\textwidth}
		\includegraphics[width=\textwidth]{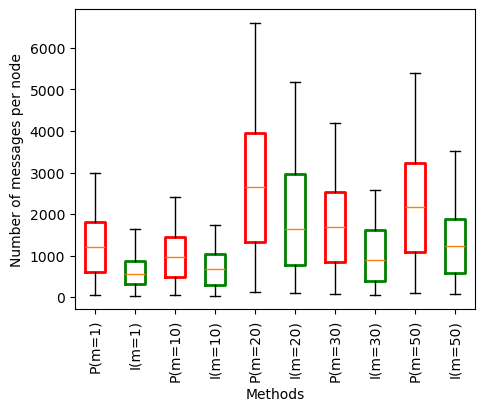}
		\caption{}
		\label{fig:boxplot_1}
	\end{subfigure}
	\begin{subfigure}[h]{0.325\textwidth}
	\includegraphics[width=\textwidth]{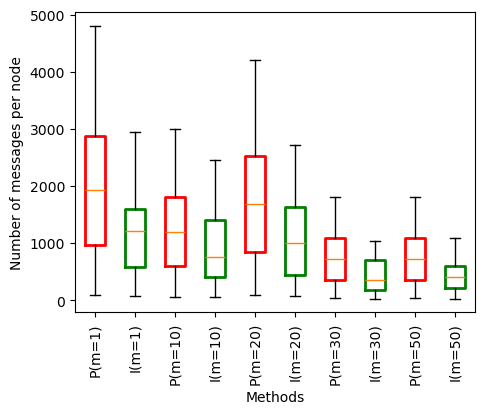}
	\caption{}
	\label{fig:boxplot_2}
 \end{subfigure}
 	\begin{subfigure}[h]{0.325\textwidth}
	\includegraphics[width=\textwidth]{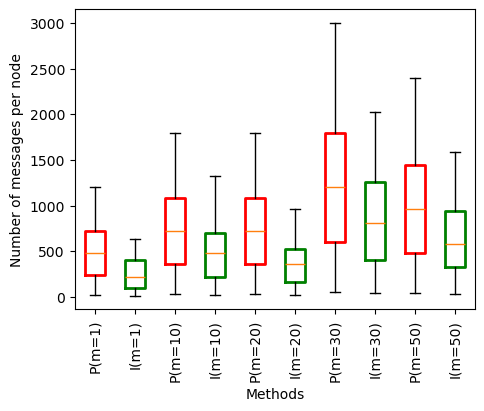}
	\caption{}
	\label{fig:boxplot_3}
\end{subfigure}
	\caption[\textit{Numbers of messages per node using 3 randomly selected graphs with $D=12$. On the horizontal axis P and I denote the original pruning method and our enhancement respectively.}]{\textit{{Numbers of messages per node using 3 randomly selected graphs with $D=12$. On the horizontal axis P and I denote the original pruning method and our enhancement respectively.  \textbf{(\ref{fig:boxplot_1})}: First,
	\textbf{(\ref{fig:boxplot_2})}: Second and
 \textbf{(\ref{fig:boxplot_3})}: Third selected graphs.}}}
	\label{fig:approximate}
\end{figure*}

\begin{table}[h]
	\centering
	\scriptsize
		\scalebox{1}{\begin{tabular}{ |l|l|l||l|l||l|l| }
		\hline
		\textbf{Nodes}&\textbf{Edges}&\textbf{Diameter}&\textbf{I}  & \textbf{P}&\textbf{Imax}  & \textbf{Pmax}\\
		\hline
		\hline
		\multicolumn{7}{|l|}{\textbf{Three autonomous networks}}\\
		\hline
		$1486$&$3422$&$9$&\textit{7.0}&$ 8.8$&\textit{1202}&$ 1413$\\
		\hline
		$2092$&$4653$&$9$&\textit{6.3}&$ 7.9$&\textit{430}&$ 552$\\
		\hline
		$6232$&$13460$&$9$&\textit{6.1}&$ 6.9$&\textit{1456}&$ 1459$\\
		\hline
		\multicolumn{7}{|l|}{\textbf{Phenomenology collaboration network}}\\
		\hline
		$10876$&$39994$&$9$&\textit{10.1}& $ 14.3$&\textit{112}&$ 120$\\
		\hline
		\multicolumn{7}{|l|}{\textbf{Gnutella peer-to-peer network}}\\
		\hline
		$9877$&$25998$&$13$&\textit{8.2}&$ 8.3$&\textit{701}&$ 787$\\
		\hline
  		\multicolumn{7}{|l|}{\textbf{Facebook artists network}}\\
		\hline
		$50515$&$819306$&$32$&\textit{19.1}&$ 23.5$&\textit{9614}&$ 9639$\\
		\hline
	\end{tabular}}
	\caption[\textit{Results for a sample of $6$ real-world networks ($1$ phenomenology collaboration network, $1$ Gnutella peer-to-peer network,  $3$ autonomous networks and $1$ Facebook artist network) on the original pruning method and our enhancement.}]{\textit{Graph properties and the number of messages with each approach for five real-world networks (the phenomenology collaboration network, the Gnutella peer-to-peer network, three autonomous networks  and the Facebook artist network). I(Imax) and P(Pmax) denote average(maximum) number of messages for our enhancement and the original pruning methods respectively. 
		}}
	\label{tab:autonomous_graph}
\end{table}

Table \ref{tab:autonomous_graph} shows the average  and the maximum number of messages per node for six real-world networks respectively.  
The number of messages per node for each technique on three random network are contrasted in Figure \ref{fig:approximate}.
The results confirm that our enhancement is better than the original pruning method.  
\subsubsection{Quality of selected most central node}

Table \ref{tab:shortest_path_1} shows the shortest path distances between the exact most central node and approximated most central nodes using the original pruning method and our enhancement method for two random graphs. 
For each of the graphs, we randomly vary $D$ in a range of values smaller than the diameter of the graph. 
\begin{table}[h]
	\centering
	\scriptsize
	\scalebox{1}{	\begin{tabular}{ |l|c|c||c|c| }
			\hline
\textbf{$D$}&\textbf{I$_1$}&\textbf{P$_1$}&\textbf{I$_2$}&\textbf{P$_2$}\\
\hline
$2$ &6&6&6&6\\
\hline
$5$ &$11$&11&$4$&5\\
\hline
$8$ &$8$&7&$2$&2\\
\hline
$12$ &$4$&4&$4$&3\\
\hline
$14$ &$0$&0&0&0\\
\hline
$20$ &0&0&0&$0$\\
\hline
$24$ &0&0&0&0\\
\hline
$28$ &0&0&0&0\\
\hline
	\end{tabular}}
	\caption[\textit{Shortest path distances between the exact most central node and approximated most central node.}]{\textit{Shortest path distances between the exact most central node and approximated most central node using the original pruning and our enhancement. We use two random graphs, one with $120$ nodes and diameter of $32$, and another with $300$ nodes and diameter of $34$. The two approaches achieve exactly the same approximations. P$_i$ and I$_i$ indicate shortest path distances for the original pruning method and our enhancement on the $i$-th random graph respectively.}}
	\label{tab:shortest_path_1}
\end{table}

When using both the original pruning method and our enhancement for leader selection based on closeness centrality, the two methods yield the same results under identical conditions.
Furthermore, our analysis confirmed the key finding of the original work \cite{masakuna2023distributed}: the two methods generally provide good approximations to the true closeness centrality when the parameter $D$ (often related to the maximum communication distance or number of iterations) is sufficiently large (see Table \ref{tab:shortest_path_1}).
This result is critical, as it confirms that the communication improvements achieved by our enhancement do not sacrifice the underlying accuracy of the centrality estimation, particularly in the regime where the original method is known to perform well.

\subsection{Hypothesis test}
\begin{table}[h]
	\centering
	\scriptsize
	\scalebox{1.}{	\begin{tabular}{ |l|c|c| }
			\hline
\textbf{Metrics}&\textbf{$p$-value}&\textbf{effect size}\\
\hline
number of messages &$0.0625$&$0.586$\\
\hline
running times &$0.0876$&$0.437$\\
\hline
closeness centrality quality &$0.0925$&$0$\\
\hline
	\end{tabular}}
	\caption[\textit{Hypothesis testing (including p-value and effect size) was conducted to compare the results of the original pruning method with our enhancement across various metrics, using all 135 graphs.}]{\textit{Hypothesis testing (including p-value and effect size) was conducted to compare the results of the original pruning method with our enhancement across various metrics, using all 135 graphs.}}
	\label{tab:hypothesis}
\end{table}
Our empirical evaluation on $135$ graphs yielded several key insights (see Table \ref{tab:hypothesis}). We observed $p$-values larger than the conventional significance threshold for the number of messages and running times when comparing the original pruning method to our enhancement. This suggests that the performance improvement, while evident on individual instances, is not statistically significant based strictly on the $p$-value criterion.
However, the analysis of effect size provides a more nuanced picture. We found a medium effect size ($0.4 \leq e < 0.8$, following the classification by Gail and Richard \cite{effectsize}) for both the number of messages and running times. This medium effect size indicates that the central tendency (mean) of the number of messages and running times using our enhancement is meaningfully different from the original pruning method. 
More importantly, the metric for the quality of the selected central nodes yielded a $p$-value of $0$ (or near-zero). This result strongly supports our approach: it confirms that our enhancement preserves the quality of the most central nodes selected by the algorithm, demonstrating that our communication efficiency gains are achieved without diminishing the core accuracy of the centrality estimates.

%% file: conclusion.tex
\section{Conclusion}
\label{sec:conclusion}
In this paper, we have effectively explored the application of the pruning method in the context of multi-packet messaging for distributed closeness centrality computations. We also enhanced the original pruning method by reducing the number of messages exchanged between nodes. Our experiments indicate that, while multi-packet messaging is necessary for larger messages using pruning, the trade-off is greater memory usage and overhead compared to single-packet messaging, which may be preferred for smaller data transfers.  Our experiment also indicates that our enhancement has reduced the number of messages exchanged between nodes, enabling pruning to alleviate the burden of multi-packet messaging, which typically increases the number of transmitted messages, all without compromising the accuracy of closeness centrality quality.

For future work, we plan to investigate the integration of our method with other centrality measures and its application in real-world scenarios, such as mobile sensor networks and social network analysis, to validate its robustness across diverse environments.